\documentclass[reprint,twocolumn,amsmath,amssymb,aps,pra]{revtex4}
\usepackage{txfonts}
\usepackage{graphicx}
\usepackage{hyperref}
\usepackage{dcolumn}
\usepackage{bm}
\usepackage{textcomp}
\usepackage{setspace}
\usepackage{float}
\usepackage{indentfirst}
\usepackage{natbib}
\usepackage{subfigure}
\usepackage{booktabs}
\usepackage{multirow}
\usepackage{txfonts}
\usepackage{amsmath}
\usepackage{braket}
\usepackage{mathrsfs}
\usepackage{ulem}
\usepackage[titletoc]{appendix}

\allowdisplaybreaks
\makeatletter

\newcommand{\Rmnum}[1]{\expandafter\@slowromancap\romannumeral #1@}
\makeatother
\hypersetup{colorlinks=true,allcolors=blue,urlcolor=cyan,pdfstartview=Fit,breaklinks=true}
\begin{document}
	\bibliographystyle{plainnat}
	\title{ Generation and Stabilization of Bound States in the Continuum in Dissipative Floquet Optical Lattices}
	\author{Yangchun Zhao$^{1}$}
	\author{Hongzheng Wu$^{1}$}
	\author{Xinguang Li$^{1}$}
	\author{Lei Li$^{2}$}
	\author{Jinpeng Xiao$^{2}$}
	\author{Zhao-Yun Zeng$^{2}$}
	\author{Yajiang Chen$^{2}$}
	\author{Xiaobing Luo$^{1}$}
	\altaffiliation{Corresponding author: xiaobingluo2013@aliyun.com}
	\affiliation{$^{1}$Zhejiang Key Laboratory of Quantum State Control and Optical Field Manipulation, Department of Physics, Jinggangshan University, Ji'an 343009, China}
	\affiliation{$^{2}$School of Mathematics and Physics, Jinggangshan University, Ji’an 343009, China}
	\date{\today}
	
	\begin{abstract}
	This paper investigates the generation and stabilization of bound states in the continuum (BICs) in a one-dimensional dissipative Floquet lattice. We find a different mechanism for the generation of stable BICs in the open one-dimensional lattice system, which stems from a peculiar dark Floquet state, a state with zero quasi-energy and negligible population on the lossy sites. Our results reveal that the evolutionary stability of BICs resulting from the dark Floquet state can be significantly enhanced, as evidenced by their very low decay rate, by increasing the driving frequency or, counterintuitively, increasing the dissipation strength. We further demonstrate that stable dark Floquet BICs can robustly persist even in nonlinear regimes. The existence of these stable dark Floquet BICs can be attributed to the role of higher-order correction terms in the effective Floquet Hamiltonian derived via the high-frequency expansion (HFE) method. Furthermore, we demonstrate that incorporating non-Hermitian dissipation can extend the parameter regime for the existence of BICs, and the dissipation-induced BICs can lead to complete reflection of wave packets. Our findings provide theoretical support for the experimental realization of stable BICs in dissipative quantum systems.
	\end{abstract}
	
   \maketitle
	
	\section{INTRODUCTION}
Bound states in the continuum (BICs) represent a fascinating class of quantum states that remain spatially localized despite having energy embedded within the continuum spectrum of scattered states. First proposed by von Neumann and Wigner in 1929 through their seminal work \cite{Neumann1929}, BICs were originally conceived as bound electronic wave functions embedded in the continuum of propagating modes. This counterintuitive phenomenon has garnered significant scientific interest due to its unique physical characteristics and has been experimentally observed across diverse systems, far beyond its initial theoretical framework. Notably, manifestations of BICs have been identified in atomic and molecular systems \cite{2,3,4}, as well as in correlated systems described by Hubbard models \cite{5,6,7} and quantum Hall systems \cite{8}. The concept has since been extended to photonic systems, where optical BICs have emerged as a prominent research frontier \cite{9,10,11,12}. This paradigm has further expanded to encompass classical wave phenomena, demonstrating remarkable universality through experimental realizations in acoustic systems \cite{13,14,15,16}, hydrodynamic wave platforms \cite{17,18,19}, and electromagnetic metamaterials \cite{20,21,22,23,24}. Recent advances in BIC research have led to an enriched classification scheme that now includes symmetry-protected BICs \cite{25} and topologically nontrivial BICs \cite{26}, highlighting the ongoing evolution of this vibrant field.
	
Floquet theory provides a powerful framework for engineering quantum systems via time-periodic modulation, allowing for dynamic control over energy spectra and the density of states. This approach has uncovered novel phenomena, including Floquet topological phase transitions \cite{27} and discrete time crystals \cite{28,29}. The integration of Floquet mechanisms with BICs has resulted in the emergence of Floquet BICs \cite{9,30,31,32,33,34,35,36,37}, which transcend the limitations of static systems while offering unprecedented precision in manipulating BIC dynamics. By strategically designing the driving field, researchers can now control the creation, destruction, movement, and mixing of the BICs in driven systems. Recently, the concept of BICs has been extended to $\mathcal{PT}$-symmetric non-Hermitian systems \cite{38,39,40,41}. The interplay between gain-loss engineering, periodic driving, and exceptional point physics has fundamentally reshaped our understanding of BIC stability and dynamical evolution \cite{42}. Generally, dissipation is viewed as detrimental to the coherence and stability of quantum states in open systems. However, recent studies have shown that carefully engineered dissipation can, counterintuitively, enhance the stability of certain quantum states \cite{43,44}. To date, it remains unclear whether stable BICs can emerge in purely dissipative non-Hermitian systems through Floquet engineering. 
	
	In this work, we propose a mechanism for generating stable Floquet BICs in an ac-driven, lossy non-Hermitian lattice. We demonstrate that the existence of stable Floquet BICs in the open one-dimensional lattice system arises from a peculiar dark Floquet state — a state characterized by zero quasi-energy and negligible population on the lossy sites. Through their temporal evolution, we demonstrate that Floquet BICs originating from the dark Floquet state exhibit enhanced stability in preserving their profiles as the driving frequency or dissipation strength increases — a finding that defies conventional wisdom. We further confirm that the dissipative system not only supports stable dark Floquet BICs, but also ensures that their stability persists in nonlinear systems. Additionally, the system supports arbitrarily configurable multi-mode dark BICs. Using the high-frequency expansion (HFE) method, we analytically elucidate the physical mechanism underlying the existence of these dark Floquet BICs. Notably, the introduction of dissipation significantly expands the parameter regime that supports BICs compared to their Hermitian counterparts. These dissipation-induced BICs are shown to facilitate the complete quantum reflection of wave packets.
	 
	 The rest of the paper is organized as follows. In Section $\textcolor{blue}{\mathrm{II}}$, we introduce the model and reveal the existence of the novel Floquet BICs in purely dissipative systems. In Section $\textcolor{blue}{\mathrm{III}}$, we numerically demonstrate that increasing dissipation enhances the stability of BICs during the dynamical evolution. In Section $\textcolor{blue}{\mathrm{IV}}$, we numerically find the complete quantum reflection of wavepacket effects enabled by the dissipation-induced BICs. In Section $\textcolor{blue}{\mathrm{V}}$, we confirm the existence of stable nonlinear BIC modes. In Section $\textcolor{blue}{\mathrm{VI}}$, we employ the high-frequency expansion (HFE) to obtain the effective Floquet Hamiltonian, demonstrating the stable BIC modes resulting from the dark Floquet state. In Section $\textcolor{blue}{\mathrm{VII}}$,  the extension to arbitrarily programmable multi-mode dark BICs is illustrated. Finally, we summarize our findings in Section $\textcolor{blue}{\mathrm{VIII}}$.
	
	\section{MODEL}

\begin{figure*}
	\includegraphics[width=1\linewidth]{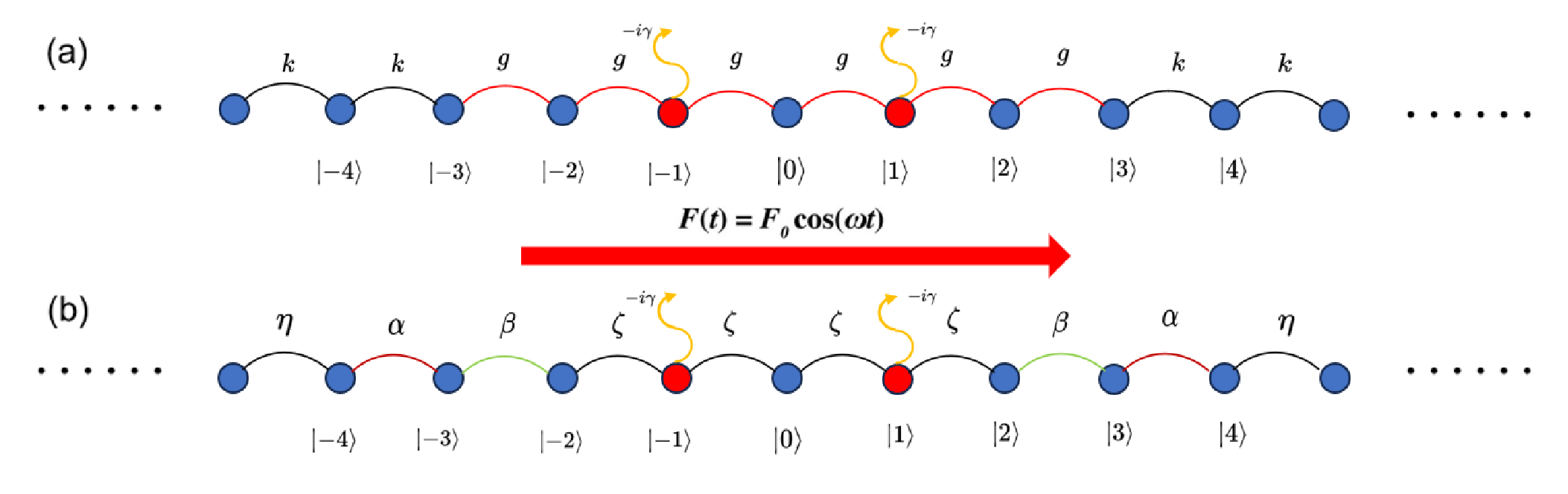}
	\caption[FIG.1]{(a) Schematic of an ac-driven tight-binding lattice with inhomogeneous nearest-neighboring hopping rates $g < k$ between lattice sites $\left|0\right\rangle,  \left|\pm1\right\rangle, \left|\pm2\right\rangle$, and $\left|\pm3\right\rangle$, with dissipative lattice sites at $\left|\pm1\right\rangle$. (b) Effective static lattice model in the high-frequency regime. The effective hopping rates $\eta$, $\zeta$, $\alpha$, and $\beta$ are given in the text by Eq. (\ref{con:21}). }
	\label{fig1}
\end{figure*}

	We consider a quantum particle hopping on a periodically driven one-dimensional optical lattice with non-Hermitian loss defects, which can be accurately described using a tight-binding Hamiltonian:
	\begin{equation}\label{con:1}
		\centering
		\begin{aligned}
			\hat{H}(t)=&\sum_{n}^{}K_{n}\left (\left|n+1\right\rangle\left\langle n \right|+\left|n\right\rangle\left\langle n+1 \right|\right)\\+&F\left(t\right)a\sum_{n}^{} n\left|n\right \rangle\left\langle n \right|-i\sum_{n}^{}\gamma_{n} \left|n\right\rangle\left\langle n \right|,\\
		\end{aligned}
	\end{equation}
where $\left| n \right\rangle$ denotes the Wannier state localized at the $n$-th lattice site ($n = 0, \pm 1, \pm 2, \ldots$), the parameter $K_n$ represents the nearest-neighbor hopping amplitude between sites $n$ and $n+1$, $a$ is the lattice constant, and $F(t) = F_0\cos(\omega t)$ describes a time-periodic driving force with amplitude $F_0$ and angular frequency $\omega$. The dissipation rate at site $n$ is characterized by $\gamma_n$. This model's dissipationless regime ($\gamma _{n}=0$) has been successfully realized in multiple experimental platforms. Prominent implementations include: coherent quantum transport of ultracold atoms in temporally modulated optical lattices through periodic phase-shaking protocols \cite{45}, and photonic Floquet engineering in femtosecond-laser-written waveguide arrays with periodic curvature modulation \cite{46,47}. In our model, we assume the lattice exhibits localized dissipation with $\gamma_{n}=\gamma$ specifically at sites $n=-1$ and $n=1$, while $\gamma_{n}=0$ at all other sites. Furthermore, we implement position-dependent hopping amplitudes where the hopping rate satisfies $K_{n}=k$ for $n\neq-3,-2,-1,0,1,2$, and $K_{-3}=K_{-2}=K_{-1}=K_{0}=K_{1}=K_{2}=g$, as illustrated in Fig. \ref{fig1}(a). Under the influence of periodic forcing, the eigenstates $\left | \psi(t) \right \rangle$ of the Hamiltonian are Floquet states, expressed as $\left | \psi(t) \right \rangle = e^{-i\varepsilon t} \left | u(t) \right \rangle$, where $\varepsilon$ represents the quasi-energy and $\left | u(t) \right \rangle$ is the Floquet mode. These states satisfy the condition $\left | u(t+T) \right \rangle = \left | u(t) \right \rangle$, where $T = 2\pi/\omega$ is the driving period. While all parameters ($F_0$, $\omega$, $K_n$, $\gamma_n$) are of unit of energy, we treat them as dimensionless in subsequent analysis because what matters are the ratios between these parameters, not their absolute values. 

 For a uniform ac-driven lattice described by the Hamiltonian $(\ref{con:1})$ with \( K_n = k \) and \( \gamma_n = 0 \)  for all \( n \), it is well known that when the ratio of the field amplitude to frequency matches a zero of the Bessel function, quasi-energy bands collapse, leading to the phenomenon of dynamical localization (DL) \cite{51}. This work demonstrates how controlled dissipation induces robust localized modes in periodically driven quantum lattices. We identify the quasi-energy of the bound eigenstates embedded within the scattering continuous spectrum-Floquet BICs-which are stabilized through the joint interplay of dissipative confinement and periodic driving. Let us now discuss the existence of Floquet BICs in the non-Hermitian dissipative  quantum lattices. To this end, let us expand the state vector \(\left | \psi(t) \right \rangle\) of the system as
	\begin{equation}\label{con:2}
\left | \psi(t)  \right \rangle  =\sum_{n}^{} C_{n} (t)\left |n \right \rangle,
	 \end{equation}
	  where the amplitudes $C_{n}(t)$ define the probabilities of
	  finding the quantum particle at lattice site $n$. The equations governing the time-dependent probability amplitudes \( C_n(t) \) are derived by substituting the decomposition (\ref{con:2}) into the time-dependent Schr\"{o}dinger equation $  
	  i\partial_t |\psi(t)\rangle = \hat{H}(t)|\psi(t)\rangle $.  
	  This procedure yields the coupled differential equations:
	\begin{equation}
		\begin{split}\label{con:3}
			i\frac{\mathrm{d} }{\mathrm{d} t} C_{n} (t)=K_{n}C_{n+1} (t)+K_{n-1} C_{n-1} (t)+aF(t)nC_{n} (t)-i\gamma _{n} C_{n} (t).
		\end{split}
	\end{equation}

The quasi-energy spectrum and corresponding Floquet states are numerically obtained by diagonalizing the time-evolution operator over one driving period: $U(T,0)=\mathcal{T}\mathrm{exp}\left(-i\int_{0}^{T} \hat{H}(t)dt \right)
$, where $\mathcal{T}$ is the time-ordering operator. Once the quasi-energy spectrum and corresponding Floquet eigenstates are computed, we identify bound states – including Floquet BICs embedded within the scattering continuum and conventional bound states outside it – by analyzing the inverse participation ratio (IPR). For the \(i\)-th quasi-energy eigenstate (\(1 \leq i \leq N\)), the inverse participation ratio (IPR) is defined as 
	\begin{equation}
\text{IPR}^{(i)} = \frac{\sum_{n} |C_n^{(i)}|^4}{\left( \sum_{n} |C_n^{(i)}|^2 \right)^2},  
	\end{equation}
where \(C_n^{(i)}\) denotes the probability amplitude at lattice site \(n\) for the \(i\)-th Floquet eigenstate \cite{36,37}. This dimensionless quantity can be used to distinguish the localized states  from extended scattering states. BICs exhibit anomalously high IPR values (\( \sim \mathcal{O}(1) \)) compared to delocalized scattering states (\( \sim \mathcal{O}(1/N) \), with \( N \) the system size). 

\begin{figure*}
		\includegraphics[width=1\linewidth]{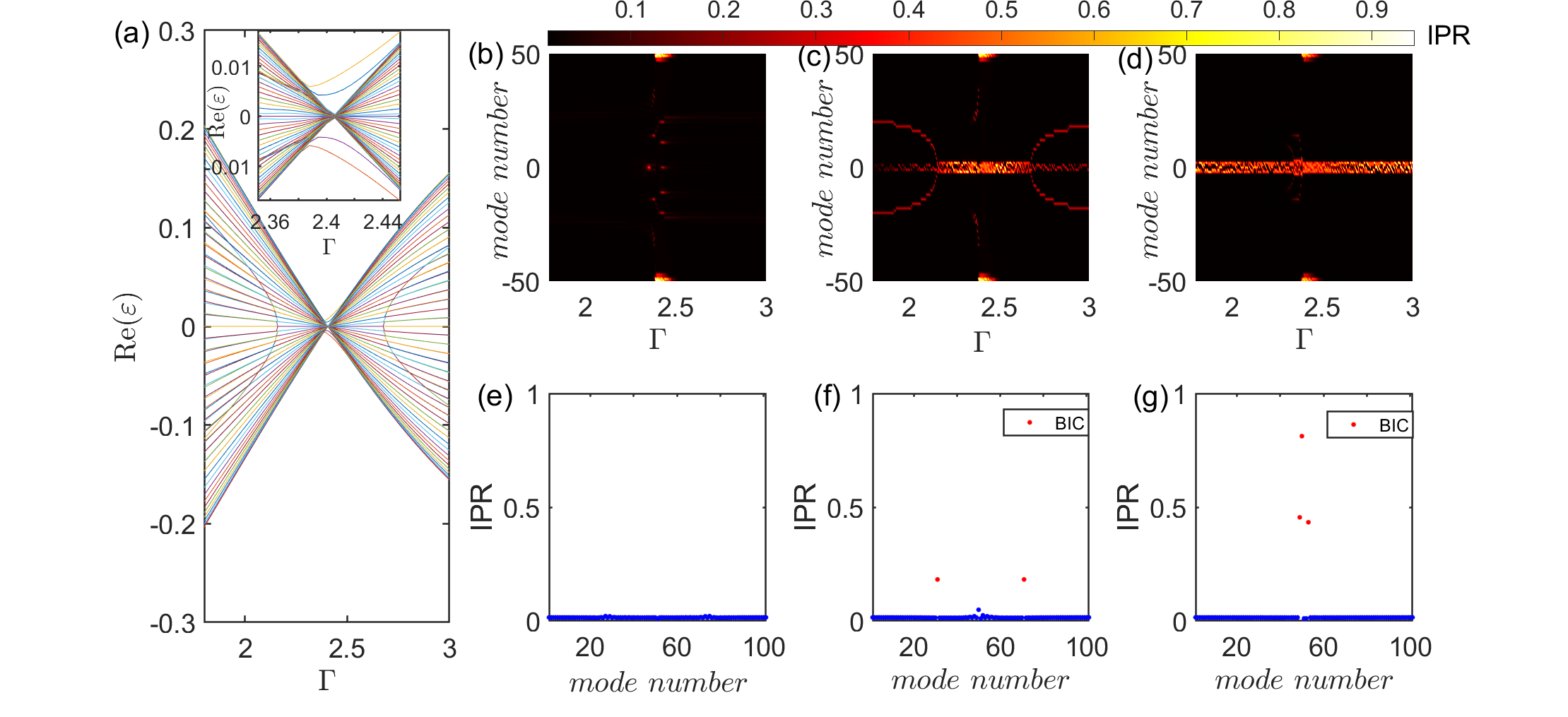}
		\caption[FIG.2]{(a) The real part of the quasi-energy spectrum $\mathrm{Re}(\varepsilon)$ vs. the normalized forcing amplitude $\Gamma = F_0 a/\omega$ at $\gamma = 0.1$. (b)-(d) Dependence of the inverse participation ratio on $\Gamma$ for the $N = 101$ Floquet eigenstates at time $t = 0$ in the ac-driven lattice of Fig. \ref{fig1}(a), for: (b) $\gamma = 0$, (c) $\gamma = 0.1$, and (d) $\gamma = 1$. The mode number on the vertical axis is ordered by increasing values of $\mathrm{Re}(\varepsilon)$. (e)-(g) The corresponding inverse participation ratio of the $N = 101$ Floquet eigenstates at $\Gamma = 1.8$ for (e) $\gamma = 0$, (f) $\gamma = 0.1$, and (g) $\gamma = 1$. The other parameters are set as $k = 0.3$, $g = 0.7k$, $a = 1$, $\omega = k/0.3$, and the total number of lattice sites $N = 101$.}
		\label{fig2}
	\end{figure*}

A representative example is shown in Fig. \ref{fig2}(a) for a lattice comprising \( N = 101 \) sites, with parameters \( k = 0.3 \), \( \omega = 1 \), $a=1$ and \( \gamma = 0.1 \). Analysis of the real part of the quasi-energy spectrum [Fig. \ref{fig2}(a)] reveals Floquet bound states outside the scattering continuum, which emerge when the dimensionless driving strength \( \Gamma = \frac{F_0a}{\omega} \) lies within the range \( 2.3 \leq \Gamma \leq 2.5 \). These states manifest as isolated spectral curves detaching from the continuum of delocalized scattering states, characteristic of bound states outside the continuum (BOCs) induced by the ac field, as discussed in Ref. \cite{37}. Among these states, two pairs of twofold-degenerate Floquet BOCs localize at the leftmost and rightmost lattice edges, a manifestation of finite-size boundary effects that vanish under periodic boundary conditions. Two additional BOC pairs (twofold-degenerate, as seen in Fig. \ref{fig3}) arise from the disparity in coupling strengths  (\(k \neq g\)) in the intermediate lattice region, which localize near lattice sites \(n = -3\) and \(n = 3\). This generation of BOCs arises from the formation of virtual defects at boundary interfaces induced by periodic driving \cite{34,37}. While the observed Floquet BOCs due to boundary effects  are well understood, our focus lies on the more elusive Floquet BICs. Figures \ref{fig2}(b)-(d) display the numerically computed IPR for all \(N\) Floquet modes as a function of the normalized driving strength \(\Gamma\), with panels (b)-(d) corresponding to dissipation coefficients \(\gamma = 0\), \(0.1\), and \(1\), respectively. The eigenmode number of Floquet states is indexed in ascending order of their real quasi-energy values, following the standard spectral ordering convention for driven quantum systems. The analysis of IPR reveals that in the dissipationless regime (\(\gamma = 0\)), BICs with nonzero \(\text{IPR}\) emerge only near \(\Gamma = 2.405\) (which corresponds to the DL condition originally revealed in Ref. \cite{48}), as evidenced by high-IPR bright points in the intermediate mode number range [see Fig.~\ref{fig2}(b)]. As the dissipation parameter \(\gamma\) increases, we observe Floquet BICs emerging across a broad range of the normalized driving strength \(\Gamma\). This dissipation-induced broadening of the BIC regime, indicated by bright high-IPR values corresponding to intermediate mode number, is clearly demonstrated in Figs.~\ref{fig2}(c)-(d). Here, even far from the DL condition, BICs persist due to the interplay of non-Hermitian dissipation and periodic driving. We also observe that near the DL conditions (\(\Gamma \approx 2.405\)), Floquet BOCs emerge as pronounced high-IPR regions (bright stripes) at the upper and lower edges of the IPR diagram in Figs.~\ref{fig2}(b)-(d), which are the same as the defect-free surface states discovered in Ref. \cite{34}. Next, we further demonstrate that the emergence of these unexpected BICs is directly mediated by finite dissipation \(\gamma\). As shown in Figs.~\ref{fig2}(e)-(g), which display the IPR for all Floquet eigenstates at a fixed driving strength of \(\Gamma = 1.8\), the system hosts 0, 2 and 3 BIC states (marked with the red dots) for dissipation strengths of \(\gamma =0\), \(0.1\) and \(1\), respectively. The synergistic interplay between non-Hermitian dissipation and periodic driving establishes a novel mechanism for generating BICs, offering an unconventional pathway to engineer embedded states in driven-dissipative quantum systems. 
	
	\section{Increasing dissipation can enhance the stability of BICs}
\begin{figure}
		\centering
		\includegraphics[width=1\linewidth]{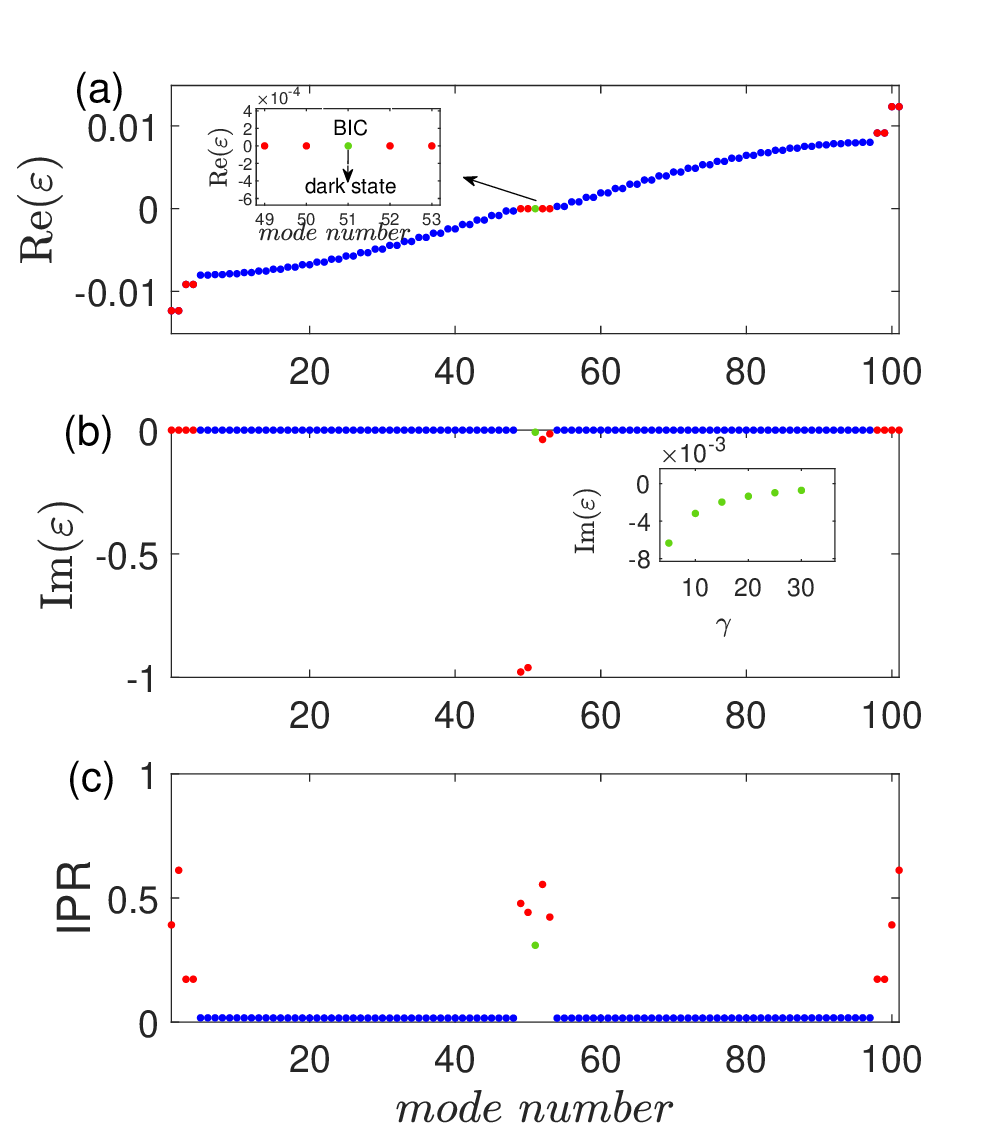}
		\caption[FIG.3]{(a) $\mathrm{Re}(\varepsilon)$ vs. mode number. The inset shows five BICs, with the green dot marking the dark Floquet BIC with nearly-zero quasi-energy. (b) $\mathrm{Im}(\varepsilon)$ vs. mode number. The inset shows the $\mathrm{Im}(\varepsilon)$ of the dark Floquet BIC approaching zero as the dissipation strength $\gamma$ increases. (c) Inverse participation ratio (IPR) vs. mode number for the $N = 101$ Floquet eigenstates at time $t = 0$ in the ac-driven lattice of Fig. \ref{fig1}(a). The other parameters are set as $k = 0.3$, $g = 0.7k$, $a = 1$, $\omega = k/0.3$, $\Gamma = 2.4308$, $\gamma = 1$, and the total number of lattice sites $N = 101$.}
		\label{fig3}
	\end{figure}

	In this section, we  analyze the stability of the BICs in the purely dissipative quantum systems. To gain a deeper understanding of the nature of BICs and their stability near the DL condition, specifically at $\Gamma=2.4308$, we have computed the real part of the quasi-energy $\mathrm{Re}(\varepsilon)$, the imaginary part of the quasi-energy $\mathrm{Im}(\varepsilon)$, and the IPR of the $N$ quasi-energy eigenstates, as shown in Figs.\ \ref{fig3}(a)-(c). We observe five BICs, coexisting with eight BOCs at the spectral edges. Notably, these BICs exhibit zero real part of the quasi-energies [see inset of Fig.\ \ref{fig3}(a)] and are characterized by non-vanishing IPR values [Fig.\ \ref{fig3}(c)], confirming their spatial localization despite residing within the continuum. Among the five BICs, one exhibits a quasi-energy imaginary part $\mathrm{Im}(\varepsilon)$ approaching zero [marked with the green dot in Fig.\ \ref{fig3}(b)], while the remaining four display  negative $\mathrm{Im}(\varepsilon)$ values. All Floquet BICs with negative imaginary quasi-energies ($\mathrm{Im}(\varepsilon)< 0$) inevitably undergo exponential temporal decay, whereas only the Floquet BIC with strictly vanishing imaginary quasi-energies ($\mathrm{Im}(\varepsilon) = 0$) persists as a dynamically stable state, immune to dissipative collapse. We further discover that the BIC with a near-zero imaginary quasi-energy exhibits enhanced stabilization under two distinct tuning protocols: (i) Its imaginary quasi-energy $\mathrm{Im}(\varepsilon)$ asymptotically approaches zero with increasing dissipation strength $\gamma$ [see the inset of Fig. \ref{fig3}(b)], and (ii) at a fixed $\gamma$, the $\mathrm{Im}(\varepsilon)$ converges toward zero as the driving frequency $\omega$ is amplified (not shown here). This BIC state with a quasi-energy of zero is in fact a dark Floquet state \cite{55, 54}, which behaves in an analogous way to the well-known dark states found in non-driven systems \cite{53}. The mechanism for generating this dark Floquet state and the stability of the associated BIC will be discussed in Sec. $\textcolor{blue}{\mathrm{VI}}$. 
	
\begin{figure}
		\centering
		\includegraphics[width=1\linewidth]{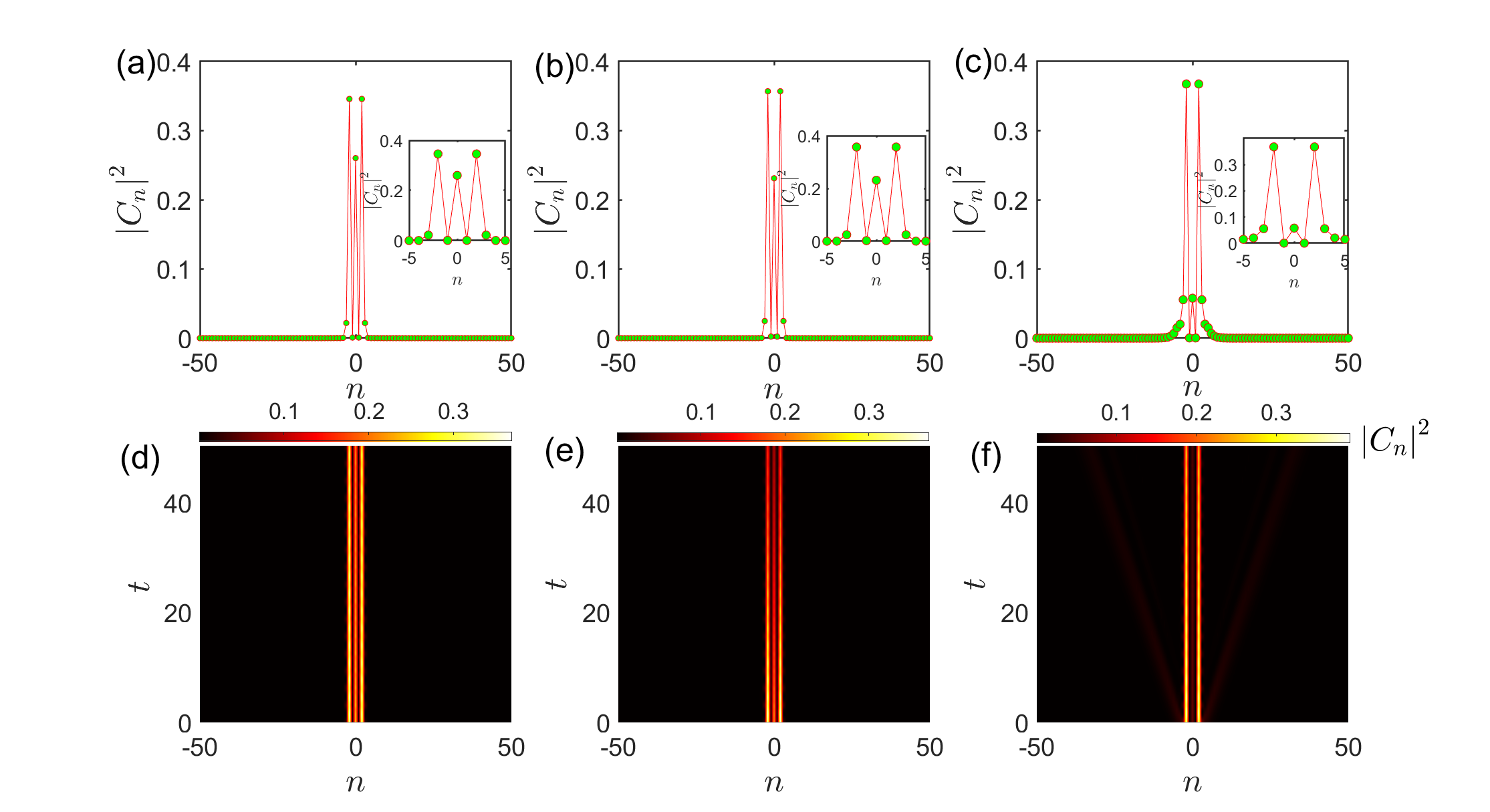}
		\caption[FIG.4]{(a)-(c) Density profile of the dark Floquet BIC at $\Gamma = 2.4308$ for $\gamma = 0.1$, $\gamma = 1$, and $\gamma = 10$, respectively. (d)-(f) Time evolution of the system with the dark Floquet BIC as the initial state over $8T$ for $\gamma = 0.1$, $\gamma = 1$, and $\gamma = 10$, respectively. The other parameters are set as $k = 0.3$, $g = 0.7k$, $a = 1$, $\omega = k/0.3$, and the total number of lattice sites $N = 101$.}
		\label{fig4}
	\end{figure}

	To intuitively illustrate how increasing dissipation strength enhances the stability of dark Floquet BICs with near-zero quasi-energy, we present in Figs.\ \ref{fig4}(a)-(f) the profiles of Floquet eigenstates under dissipation strengths $\gamma=0.1$, $1$, and $10$, along with their temporal evolution over 8 driving periods ($8T$). Analysis of the spatial profiles reveals that these dark Floquet BICs show negligible occupation at the leaky sites ($n = \pm 1$). In the presence of relatively weak dissipation, the system can be effectively modeled within a five-state reduced Hilbert space, localized at $n = -2, -1, 0, 1, 2$, and exhibits suppressed coupling to other sites. The dynamical robustness of these dark Floquet BICs in the lossy lattice is demonstrated in Figs. \ref{fig4}(d)-(f), where their stable propagation persists for over 8 driving cycles with little population decay, despite varying dissipation strengths. Furthermore, during the evolution, the dark Floquet BIC for $\gamma=10$ displays brighter and more stable signatures compared to that for $\gamma=1$, which is consistent with the trend of Im($\varepsilon$) approaching zero as observed in the inset of Fig.\ \ref{fig3}(b). This stabilization trend illustrates a dissipation-engineered  stabilization mechanism for the Floquet BICs in the lossy non-Hermitian lattice system.
	
	To establish a quantitative framework for assessing the dynamical stability of these dark Floquet BICs, we will subsequently introduce the physical quantity of decay rate to characterize their population leakage over time. Given that the non-Hermitian Hamiltonian \;(\ref{con:1}) is purely dissipative, the norm of the quantum system's state $\left|\psi(t)\right\rangle$ exhibits decay dynamics governed by \cite{49}: 
	
	\begin{equation}\label{con:5}
		\frac{\mathrm{d}}{\mathrm{d} t} \left\langle\psi(t)|\psi(t)\right\rangle =i\left\langle\psi(t)\right|\hat{H}^{\dagger }-\hat{H}\left|\psi(t)\right\rangle=-2\sum_{n}^{}\gamma _{n}\left| C_{n} (t)\right|^{2}.
	\end{equation}
	
	\begin{figure}
		\centering
		\includegraphics[width=1\linewidth]{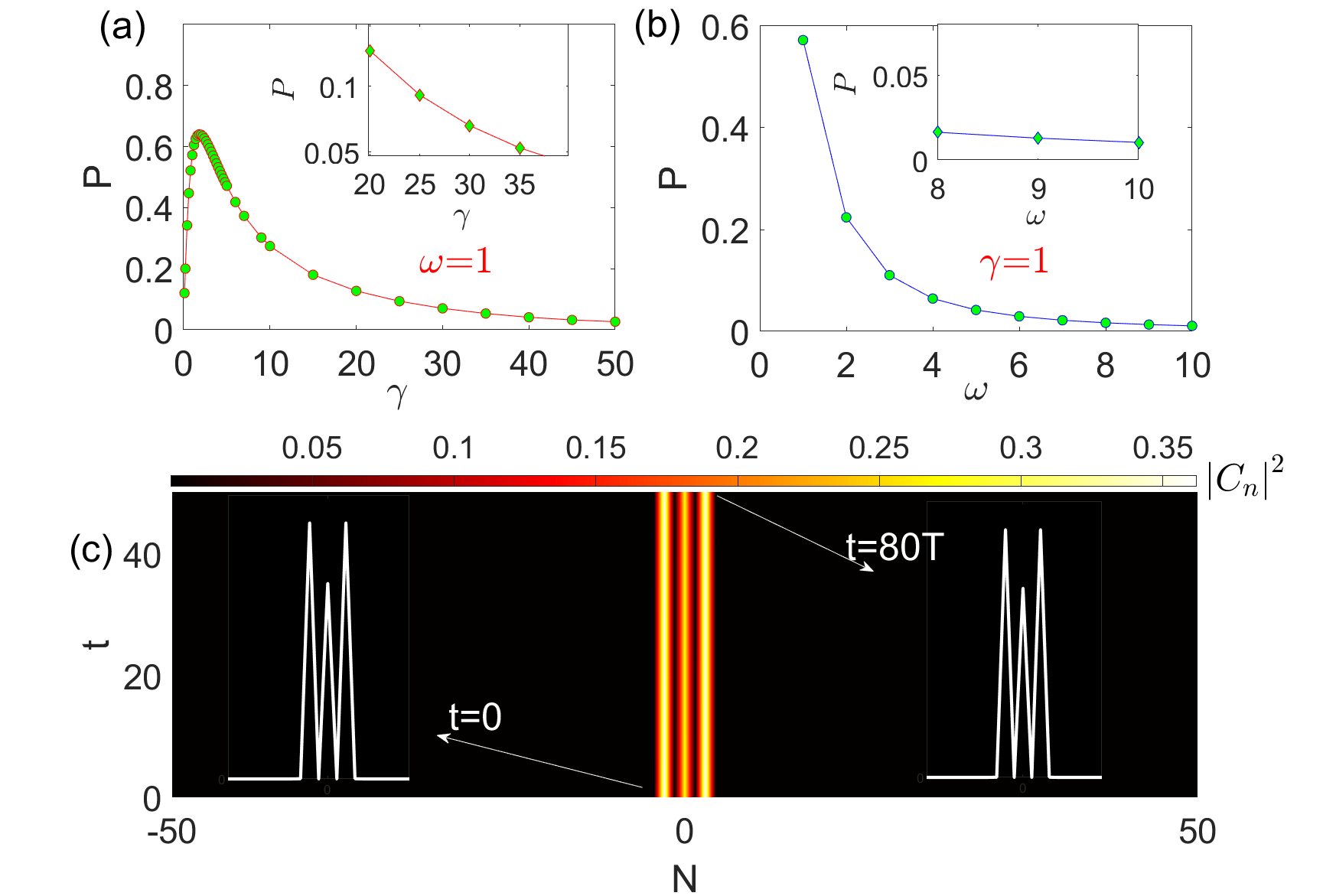}
		\caption[FIG.5]{(a) Decay probability $P$ at $t=8T$ for a quantum particle initialized in the dark Floquet BIC state, plotted as a function of dissipation strength $\gamma$ with $\Gamma=2.4308$ and $\omega=k/0.3$. (b) Driving-frequency-dependent decay probability $P$ at $\gamma=1$, measured at $t=50$ evolution. (c) Long-term stability of the dark Floquet BIC over $80T$ evolution at $\Gamma=2.40509$, $\omega=10$, and $\gamma=1$. Inset: profiles of the dark Floquet BIC at initialization ($t=0$) and terminal evolution stage ($t=80T$). Other parameters: $k=0.3$, $g=0.7k$, $a=1$, $N=101$.}
		\label{fig5}
	\end{figure}
	
	Assuming the quantum particle is initialized in the dark Floquet BIC mode at \( t = 0 \), its subsequent dynamics for \( t > 0 \) are governed by the equations of motion (\ref{con:3}). Due to the existence of pure loss in Hamiltonian (\ref{con:1}), whenever the quantum particle visits the sites of $\pm 1$, it will leak out at a certain rate according to Eq. (\ref{con:5}). Given the ability to detect the position of the site from which the probability of the quantum particle leaks out, one can obtain the total decay probability $P$ for all leaky sites. According to Eq. (\ref{con:5}), we have
	\begin{equation}\label{con:6}
		\begin{split}
			P =2\sum_{n}^{} \gamma_{n}  \int_{0}^{t}\left |C _{n} \left ( t \right ) \right |^{2} \mathrm{d}t.
		\end{split}
	\end{equation}
To investigate the decay characteristics of dark Floquet BICs, we numerically solved Eq.\ (\ref{con:3}) under open boundary conditions and recorded the total decay probability $P$ at $t = 50$. As shown in Fig.\ \ref{fig5}(a) with \( \omega \) fixed at 1, the decay rate initially increases with the dissipation coefficient \( \gamma \) in the weak dissipation regime, reaches a peak at around \( \gamma \approx 2 \), and then, as \( \gamma \) continues to increase, the decay rate drops to a low value of approximately \( \sim 0.05 \) (refer to the inset). To further confirm the relationship between the stability of dark BICs and the driving frequency, we examined the decay rate as a function of frequency, with the dissipation coefficient $\gamma$ fixed at 1. As depicted in Fig.\ \ref{fig5}(b), the decay rate exhibits a decreasing trend with increasing frequency, indicating that the quantum dynamics of the dark Floquet BIC becomes more stable at higher frequencies. For instance, the decay rate $P$  at $ t=50$  reaches an ultralow value of 0.0107 when \(\omega=10\). This dependence indicates that increasing the dissipation strength, or increasing the driving frequency, can enhance the dynamical stability of dark Floquet BICs. As an example, in Fig.\ \ref{fig5}(c), we show the evolution of the dark Floquet BIC mode over a time of $80T$ at  $\gamma=1$ and $\omega=10$, where we observe that the profile of the initial and final states exhibits almost no change, with population solely at $n=-2$, $0$, and $2$.
	
	\section{Increasing dissipation can enhance the reflection effect}
\begin{figure}
		\centering
		\includegraphics[width=1\linewidth]{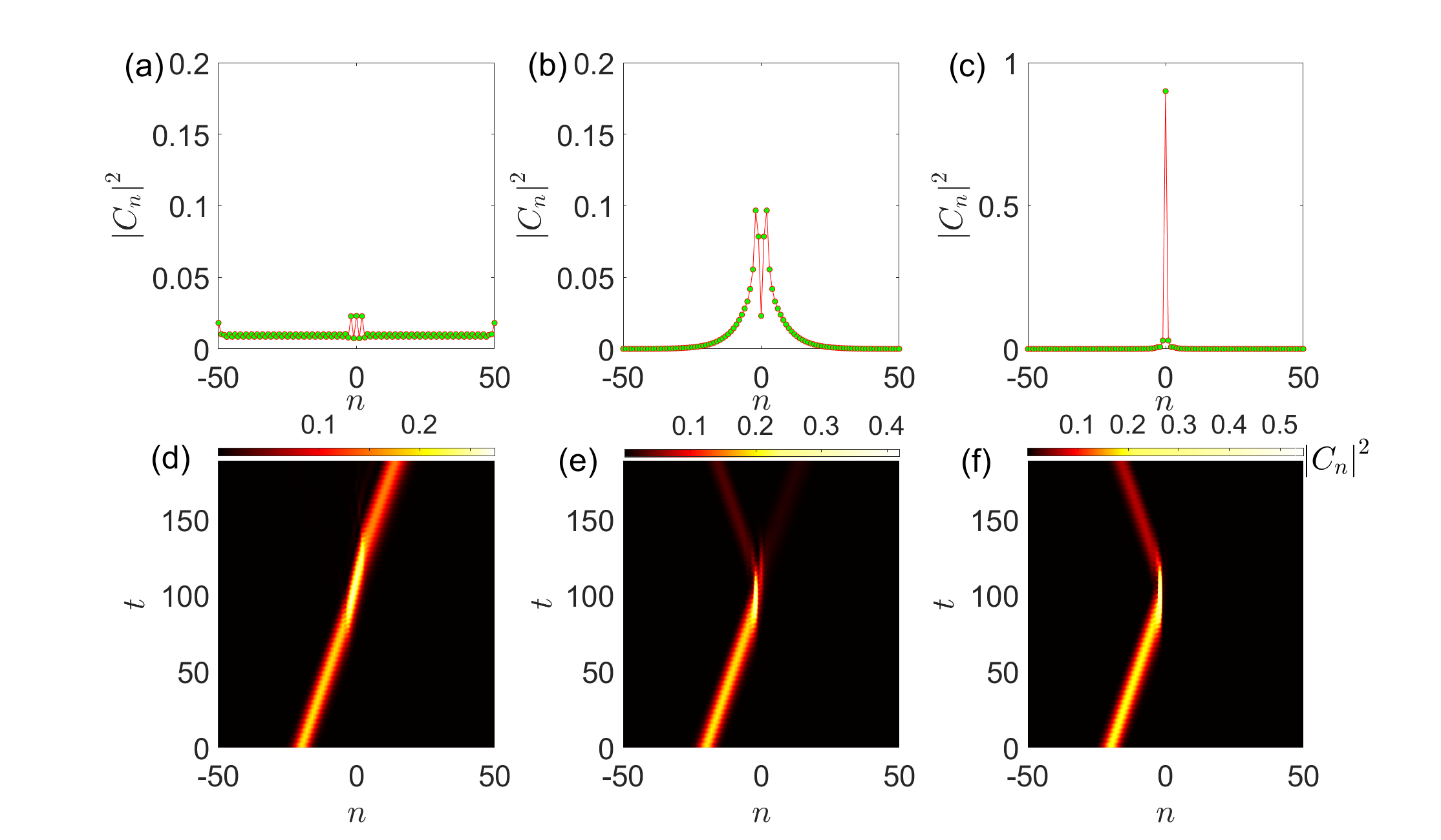}
		\caption[FIG.6]{(a)-(c) Spatial profiles of Floquet BICs at $\Gamma = 1.8$ with dissipation $\gamma = 0$ (Hermitian limit), $0.1$ (b), and $1$ (c). (d)-(f)  Wavepacket evolution over $30T$  for (d) $\gamma = 0$, (e) $\gamma = 0.1$, and (f) $\gamma = 1$. The wavepacket is initialized as:  $\left\langle n|\psi(0) \right \rangle=\mathrm{exp}\left [-(n- n_{0})^{2}/\omega _{0} ^{2} -i\pi n/2 \right ]$ with $\omega_0 = 4$, $n_0 = -20$. Other parameters: $k = 0.3$, $g = 0.7k$, $a = 1$, $\omega = k/0.3$, $N = 101$.}
		\label{fig6}
	\end{figure}

In this section, we investigate the quantum scattering of a wave packet in the ac-driven lattice of Fig.\ \ref{fig1}(a) with the normalized driving strength $\Gamma = 1.8$. As demonstrated in Figs.\ \ref{fig2}(e)–(g), at $\Gamma = 1.8$, the introduction of dissipation leads to the emergence of new BICs within the system. Figures\ \ref{fig6}(d)-(g) illustrate the spatiotemporal evolution of a Gaussian wave packet on the ac-driven lattice with different dissipation coefficients $\gamma = 0, 0.1, 1$. The other system parameters are set to \( k/\omega = 0.3 \), \(\omega = 1\), and \(\gamma = 1.8\). The initial Gaussian wave packet is defined as $\left\langle n|\psi(0)\right\rangle=\mathrm{exp}\left[-(n- n_{0})^{2}/\omega _{0} ^{2} -i\pi n/2 \right]$, with momentum \( p = \pi/2 \), centered at \( n_0 = -20 \), and width \( \omega_0 = 4 \). As shown in Fig.\ \ref{fig6}(d), the Gaussian wave packet propagates through the lattice without reflection at \(\gamma = 0\). When dissipation is introduced (\(\gamma > 0\)), the wave packet experiences partial reflection from the intermediate defective region of the lattice [Fig.\ \ref{fig6}(e)]. Remarkably, at \(\gamma = 1\) [Fig.\ \ref{fig6}(f)], the reflection becomes complete, demonstrating a dissipation-induced transition from transmission to total reflection. The mechanism behind this phenomenon can be revealed by inspecting the BIC density profiles across distinct dissipation strengths, as shown in Figs.\ \ref{fig6}(a)–(c). For example, when the dissipation strength is absent ($\gamma=0$), no BICs are observed in the system, and all eigenstates remain spatially extended, as illustrated in Fig.\ \ref{fig6}(a). As a result, the quantum wave packet can fully transmit  through the lattice. As the dissipation in the system increases, the BICs emerge and become more localized in the intermediate region [see Figs.\ \ref{fig6}(c)-(d)]. Therefore, the wave packet will be reflected by this localized BIC. These observations reveal that the emergence of dissipation-induced Floquet BICs enables complete transmission-to-reflection transitions of wavepacket. Crucially, stronger dissipation enhances the spatial localization of BICs, which progressively suppresses wavepacket transmission through the intermediate lattice, ultimately leading to complete reflection. The emergence of these BICs in non-Hermitian, periodically driven lattices enables unprecedented possibilities for steering wavepacket propagation and engineering dissipation dynamics, offering a novel paradigm for nontopologically robust transport control in quantum systems.
	
\begin{figure}
		\centering
		\includegraphics[width=1\linewidth]{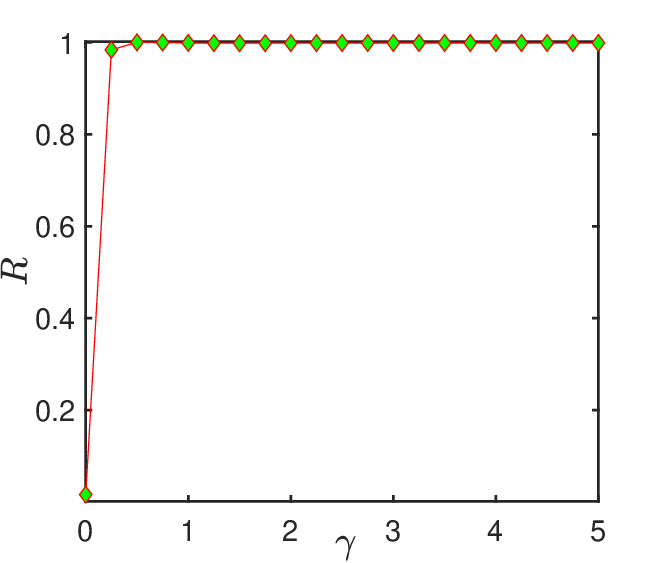}
		\caption[FIG.7]{Reflectivity versus the dissipation strength $\gamma$ at $\Gamma$=1.8. The initial condition and other parameters are the same as in Fig.  \ref{fig6}. }
		\label{fig7}
	\end{figure}

To enhance the intuitive grasp of the transmission-to-reflection transition, we introduce the reflectivity \( R \), which is defined as
	
	\begin{equation}
		\begin{split}
			R=\frac{\sum_{n=-50}^{0}\left |C_{n}(t_f)\right |^{2}}{\mathcal{N}},
		\end{split}
	\end{equation}
     where the norm is defined by $\mathcal{N}=\sum_{n=-50}^{50}\left |C_{n}(0)\right |^{2}$ , which is calculated at $t=0$.
	As the dissipation parameter increases, the sharp transition between quantum reflection and transmission is illustrated in Fig.\ \ref{fig7}, where we plot the reflectivity \( R \) at \( t_f = 30T \) as a function of \( \gamma \), using the same initial state and system parameters as in Fig.\ \ref{fig6}. We observe that at large dissipation strengths (e.g., \(\gamma = 1\)), the reflectivity \(R\) approaches unity, indicating that the quantum wave packet completely fails to transmit through the central site \(|0\rangle\) and exhibits no decay. Notably, despite this behavior of no decay, we observe in Fig. \ref{fig6}(c) a change in the wave packet brightness before and after reflection, suggesting deformation and excitation of the wave packet after reflection.
	
	\section{Nonlinearity does not affect the stability of BIC}
	In the preceding sections, we have demonstrated the existence of stable dark Floquet BIC modes in linear systems. However, a natural question arises: Can these dark Floquet BICs persist in nonlinear systems and propagate stably, or do nonlinear interactions destabilize them? 
	
\begin{figure}
		\centering
		\includegraphics[width=1\linewidth]{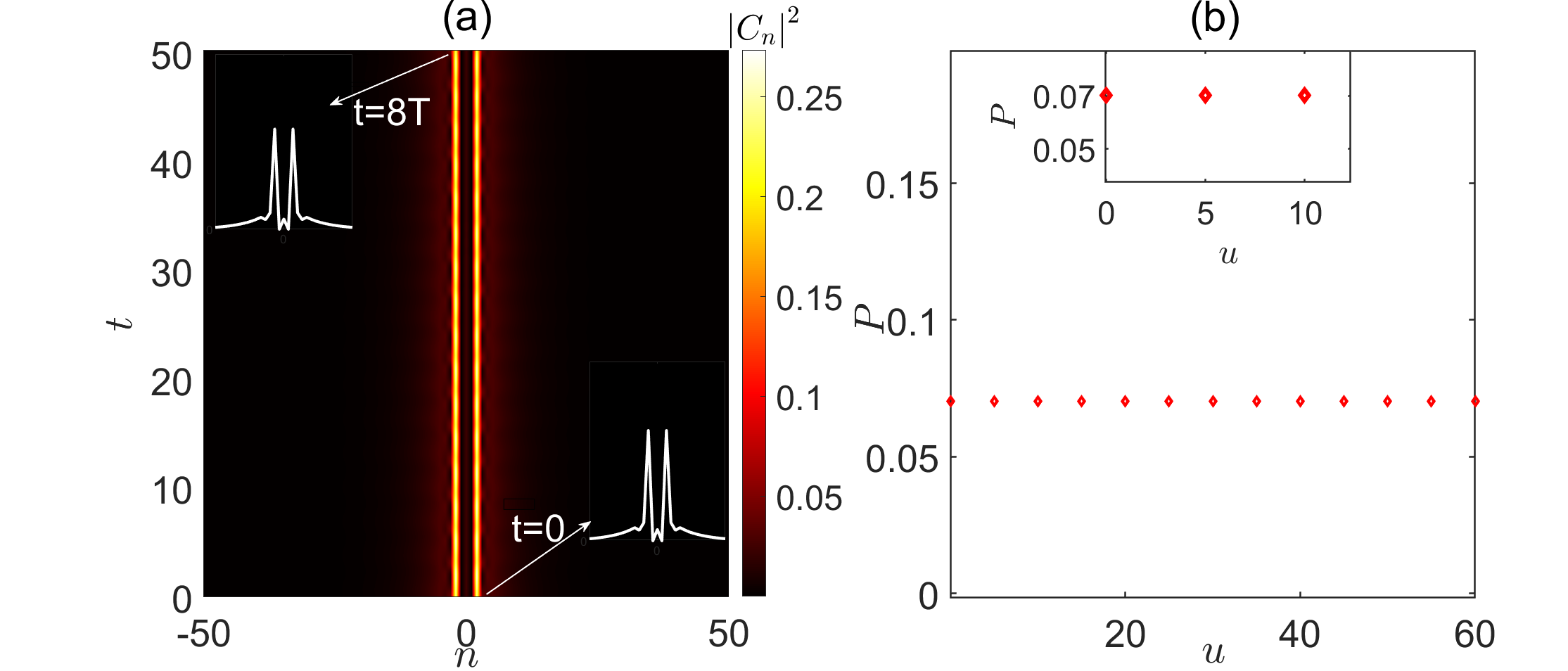}
		\caption[FIG.8]{(a) Nonlinear quantum dynamics of the dark Floquet BIC in a dissipative lattice model $(\ref{con:8})$ with $U_{-1}=U_1=u=1$, $U_{n}=0$ for $n\neq\pm 1$, and $\gamma=30$. (b) Decay rate $P$ measured at $t=8T$ versus nonlinearity strength $u$.  Inset in (a): profiles at $t=0$ (initial) and $t=8T$ (final). Other parameters: $k=0.3$, $g=0.7k$, $a=1$, $\omega=k/0.3$, and $N=101$.}
		\label{fig8}
	\end{figure}

	The equations of motion for the probability amplitudes in the nonlinear model are given by: 
	\begin{equation}\label{con:8}
		\begin{split}
			i\frac{\mathrm{d} }{\mathrm{d} t} C_{n} (t)&=K_{n}C_{n+1} (t)+K_{n-1} C_{n-1} (t)+aF(t)nC_{n} (t)\\
			&-i\gamma _{n} C_{n} (t)- U_{n}\left | C_{n}(t)\right |^{2}C_{n} (t). 
		\end{split}
	\end{equation}
	In Fig.\ \ref{fig8}(a), we numerically simulate the spatiotemporal evolution of the system under a nonlinear interaction localized at sites \(n = \pm 1\) (where \(U_n = u\) for \(n = \pm 1\) and \(U_n = 0\) otherwise), with the parameters \(u = 1\), \(\gamma = 30\), while keeping other parameters identical to those in Fig.\ \ref{fig3}. The initial state is the dark BIC obtained in the linear case (\(u = 0\)), with other parameters being the same as those in Fig.\ \ref{fig8}(a). We observe that this BIC propagates stably in the designed nonlinear system, with its shape at \(t = 0\) nearly identical to that at \(t = 8T\). Additionally, we compute the decay probabilities \( P \) at \( t = 8T \) and find that these values are extremely small and exhibit negligible dependence on the nonlinear interaction strength \( u \), as illustrated in Fig.\ \ref{fig8}(b). These findings demonstrate the existence of stable nonlinear dark Floquet BICs, suggesting promising applications in systems such as Bose-Einstein condensates and nonlinear optical waveguides.  

	\section{Methods}
	In this section, we will provide analytical insights into the existence of dark Floquet BICs. In our model, we are interested in the high-frequency limit where $K_{n}  \ll \omega$, regardless of whether the strength of the drive is weak or strong. In this scenario, it is convenient to make a transformation into a rotating frame, which avoids the need to calculate an infinite sum of series in the high-frequency expansion (HFE). The time-periodic unitary rotation operator is defined as
		\begin{equation}
			\hat{S}(t)=\mathrm{exp}\bigg[ {-i\int_{0}^{t}dtF\left(t\right)a\sum_{n}^{} n\left|n\right \rangle\left\langle n \right|}\bigg].
		\end{equation}
    In the rotating frame the model Hamiltonian takes the form
	\begin{equation}\label{con:9}
		\begin{split}
			\hat{{H}'}(t)&=\hat{S}^{\dagger } (t) \hat{H}(t) \hat{S} (t)-i\hat{S}^{\dagger}(t)\frac{\mathrm{d}\hat{S}(t)  }{\mathrm{d} t}\\
			&=\sum_{n}^{}K_{n} \left (e^{-i\Gamma sin(\omega t)}\left|n+1\right\rangle\left\langle n \right|+e^{i\Gamma sin(\omega t)}\left|n\right\rangle\left\langle n+1 \right|\right)\\&-i\gamma_{n}\sum_{n}^{}\left|n\right\rangle\left\langle n \right|,
		\end{split}
	\end{equation}
	where we defined $\Gamma =\frac{aF_{0} }{\omega }$. The Hamiltonian in the rotating frame can be  expanded in its Fourier harmonics as
		\begin{equation}
		\begin{split}
			\hat{{H}' }(t)=\sum_{l=-\infty}^{\infty} e^{il\omega t} \hat{{H}' }_{l}.
		\end{split}
	    \end{equation}
   Then, using the HFE in the rotating frame, the effective (time-independent) Floquet Hamiltonian can be expressed in terms of a series expansion in powers of $1/\omega$ \cite{50,51,52},
   	\begin{equation}
   	\begin{split}
   		\hat{H}_\text{{eff}}=\sum_{n=0}^{\infty} \frac{1}{\omega ^{n}}\hat{H}_{\mathrm{eff}}^{(n)}. 
   	\end{split}
   \end{equation}
   The leading-order term in the effective Hamiltonian expansion corresponds to the time-average of the driven Hamiltonian
   \begin{equation}
	\begin{split}
		\hat{H}_\text{{eff}}^{(0)}&=\hat{{H}' }_{0}=\frac{1}{T} \int_{0}^{T} dt\hat{{H}' }(t)\\
		&=\sum_{n}^{}K_{n}\left [ J_{0}(\Gamma)\left | n  \right \rangle\left \langle n+1 \right | + J_{0}(\Gamma)\left | n+1  \right \rangle\left \langle n \right|\right ]-i\gamma_{n} \left | n \right \rangle \left \langle n \right |,       
	\end{split}
   \end{equation}
    where \( J_0 \) is the zeroth-order Bessel function of the first kind. This result is derived by applying the identity
   \begin{equation}
   e^{ i\Gamma \sin \omega t} = \sum_{l=-\infty}^{\infty} J_{l}(\Gamma )  e^{il\omega t}.
   \end{equation}
   This Bessel-function-type renormalization enables precise control of the tunneling matrix element, allowing us to entirely ‘switch off' the nearest-neighbor coupling. This effect is known as dynamical localization or band collapse \cite{48}.
   
  The leading correction term in the HFE is given as 
	\begin{equation}
		\begin{split}
			\hat{H}_\text{{eff}}^{(1)}=\sum_{l=1}^{\infty } \frac{1}{l} \left [ \hat{{H}' }_{l},\hat{{H}' }_{-l}\right]=0,      
		\end{split}
	\end{equation}
with the Fourier components of the Hamiltonian in the rotating frame reading
	\begin{equation}
     \hat{{H}' }_{l} = \sum_n K_n \left[ J_l(\Gamma) | n \rangle \langle n + 1| + J_{-l}(\Gamma) | n + 1 \rangle \langle n| \right],
    \end{equation}
   for $l\neq0$. Due to \(\hat{H}_\text{{eff}}^{(1)} = 0 \), we push the perturbation analysis up to the second order of \( 1/\omega \)
	\begin{equation}
		\begin{split}
			\hat{H}_{\text{eff}}^{(2)}&=\sum_{l \neq 0} \left( \frac{[\hat{{H}' }_{-l}, [\hat{{H}'}_0, \hat{{H}' }_l]]}{2 l^2} + \sum_{l' \neq 0,l} \frac{[\hat{{H}' }_{-l'}, [\hat{{H}' }_{l'-l}, \hat{{H}' }_l]]}{3 ll'} \right)\\
			&=-Q(\Gamma){\left[ K_{n}K_{n+1}^2 - 2K_n^3 + K_{n}K_{n-1}^2 \right]},  
		\end{split}                 
	\end{equation}
	
	where $	Q(\Gamma)$ is given by 
	\begin{equation}\label{con:18}
		\begin{split}
			Q(\Gamma)=-\sum_{l,j\neq 0}\frac{1}{lj}J_l(\Gamma)J_j(\Gamma)J_{j-l}(\Gamma). 
		\end{split}
	\end{equation}
   Thus, we obtain the approximate effective (time-independent) Hamiltonian up to the second order in $1/\omega$,
	\begin{equation}\label{con:19}
		\begin{split}
			\hat{H}_{\text{eff}}&=\hat{H}_{\text{eff}}^{(0)}+\frac{1}{\omega }\hat{H}_{\text{eff}}^{(1)}+\frac{1}{\omega^2 }\hat{H}_{\text{eff}}^{(2)}\\
			&=\sum_{n}^{}\Theta_{n}[\left | n \right \rangle\left \langle n+1 \right |+\left | n+1  \right \rangle\left \langle n \right |]-i\gamma _{n} \left |n \right \rangle \left \langle n \right |,
		\end{split}
	\end{equation}
  where the effective hopping rate $\Theta_{n}$ between states  $\left | n  \right \rangle$ and $\left | n+1  \right \rangle $  is given by
	\begin{equation}\label{con:20}
		\begin{split}
			\Theta_{n}=K_{n}J_{0}(\Gamma)-\frac{Q(\Gamma)}{{\omega}^2}{\left[ K_{n}K_{n+1}^2 - 2K_n^3 + K_{n}K_{n-1}^2 \right]}.
		\end{split}
	\end{equation}

   From Eq.\ $\;(\ref{con:20})$, we observe that for a uniform lattice (where \( K_n = k \) for all \( n \)), the second-order correction (the term proportional to $1/\omega^2$) to the effective hopping rate $\Theta_{n}$ vanishes. However, when  $K_{n}$ are not all equal, near the DL condition [corresponding to $J_0(\Gamma)=0$], this second-order correction becomes non-negligible and plays a critical role in modifying the physical properties of the system.
   
	For the lattice shown in Fig. \ref{fig1}(a), i.e., for $K_{n}=k$ for $n\neq-3,-2,-1,0,1,2$, and $K_{-3}=K_{-2}=K_{-1}=K_{0}=K_{1}=K_{2}=g$, the effective static lattice, described by the Hamiltonian Eq. $\;(\ref{con:19})$, is depicted in Fig. \ref{fig1}(b). The effective hopping rates $\eta(\Gamma)$, $\zeta(\Gamma)$, $\alpha(\Gamma)$, and $\beta(\Gamma)$ are given by Eq. $\;(\ref{con:20})$ as follows:
	\begin{equation}\label{con:21}
		\begin{split}
			\eta(\Gamma)=& kJ_0(\Gamma),\\
			\zeta(\Gamma)=&gJ_{0} (\Gamma),\\
			\alpha(\Gamma)=& kJ_0(\Gamma) - \frac{Q(\Gamma)}{{\omega}^2}k\left( g^2 -k^2 \right),\\
			\beta(\Gamma)=& gJ_0(\Gamma) + \frac{Q(\Gamma)}{{\omega}^2}g\left( g^2 -k^2 \right).    
		\end{split}
	\end{equation}

By tuning the driving parameter \(\Gamma\), we can selectively nullify specific effective hopping terms. For instance, setting \(\beta(\Gamma) = 0\) eliminates the coupling between the states \(|-3\rangle \leftrightarrow |-2\rangle\) and \(|3\rangle \leftrightarrow |2\rangle\), effectively isolating a five-state subsystem (comprising states $|-2\rangle$, $|-1\rangle$, $|0\rangle$, $|1\rangle$, and $|2\rangle$) as illustrated in Fig. \ref{fig1}(b). This subsystem is governed by an effective five-state Hamiltonian:
	\begin{equation}\label{con:22}
		\begin{split}
			\hat{H}_{\text{eff}}=\begin{bmatrix}
				0 & \zeta  & 0 & 0 & 0\\
				\zeta & -i\gamma & \zeta & 0 & 0\\
				0 & \zeta & 0 & \zeta & 0\\
				0 & 0 & \zeta& -i\gamma & \zeta \\
				0 & 0 & 0 & \zeta & 0
			\end{bmatrix}.
		\end{split}
	\end{equation}
	As is known, for an \(N\)-state time-independent Hamiltonian (\(N\) odd), if the diagonal elements satisfy \(H_{ii} = 0\) for all odd \(i\), the eigenvalue equation \(H\mathbf{w} = E\mathbf{w}\) admits a zero-energy eigenstate (dark state) \cite{53}. This state is characterized by vanishing amplitudes at even-indexed states: \( \mathbf{w} = \left(w_1, 0, w_3, 0, \dots, w_N\right)^T, \) where \(w_{2k} = 0\) for \(k = 1, 2, \dots, (N-1)/2\). For $\hat{H}_{\rm{eff}}$$\;(\ref{con:22})$, there obviously exists such a dark state, and the eigenvalues of $\hat{H}_{\rm{eff}}$  $\;(\ref{con:22})$ can be analytically derived as
	\begin{equation}
		\begin{split}
			E_{1} &= 0, \\
			E_{2,3}& = -\frac{i\gamma}{2} \pm   \frac{\sqrt{12\zeta^2 -\gamma^2}}{2}, \\
			E_{4,5}&= -\frac{i\gamma}{2} \pm  \frac{\sqrt{(2\zeta  + \gamma)(2\zeta  - \gamma)}}{2}.
		\end{split}
	\end{equation}

The eigenstate corresponding to \( E_1=0 \), known as the dark state \cite{55,54,53}, is stable due to its purely real eigenvalue (zero imaginary part), corresponding to no exponential decay or growth. In contrast, other eigenstates possess eigenvalues with negative imaginary parts  (i.e., \( \text{Im}(E) < 0 \)), leading to exponential decay in their amplitudes over time. The eigenstates embedded in the continuum band correspond to the BICs identified in our previous numerical studies. Among these, the eigenstate with zero energy is specifically designated as the dark Floquet BIC. As shown previously in Figs.\ \ref{fig4}(a)-(c), we designed the driving parameter $\Gamma=2.4308$, which is near the DL condition, resulting in $\beta=0$ in Eq.\ $(\ref{con:21})$, numerically confirming the existence of a BIC with negligible populations at the lossy sites $\pm 1$, which is a distinctive feature of the Dark Floquet state \cite{55,54}. 
	
	\section{Extension to multi-mode dark BICs}
	
The dark Floquet BICs can also be generated in a more general framework by appropriately engineering the parameters \( K_n \) and \( \gamma_n \) in model  $(\ref{con:1})$ as follows:
		\begin{equation}\label{con:24}
		\begin{split}
			K_{n} =&\begin{cases}
				g,-M\le n\le M-1, M\in \mathrm {integers}, \\
				k,\mathrm{else}.
			\end{cases}\\
			\gamma _{n}=&\begin{cases}
				\gamma ,n=-M+2m,m=1,2,\cdots,M-1,\\
				0,\mathrm{else}.
			\end{cases}
		\end{split}
	\end{equation}

\begin{figure}
		\centering
		\includegraphics[width=1\linewidth]{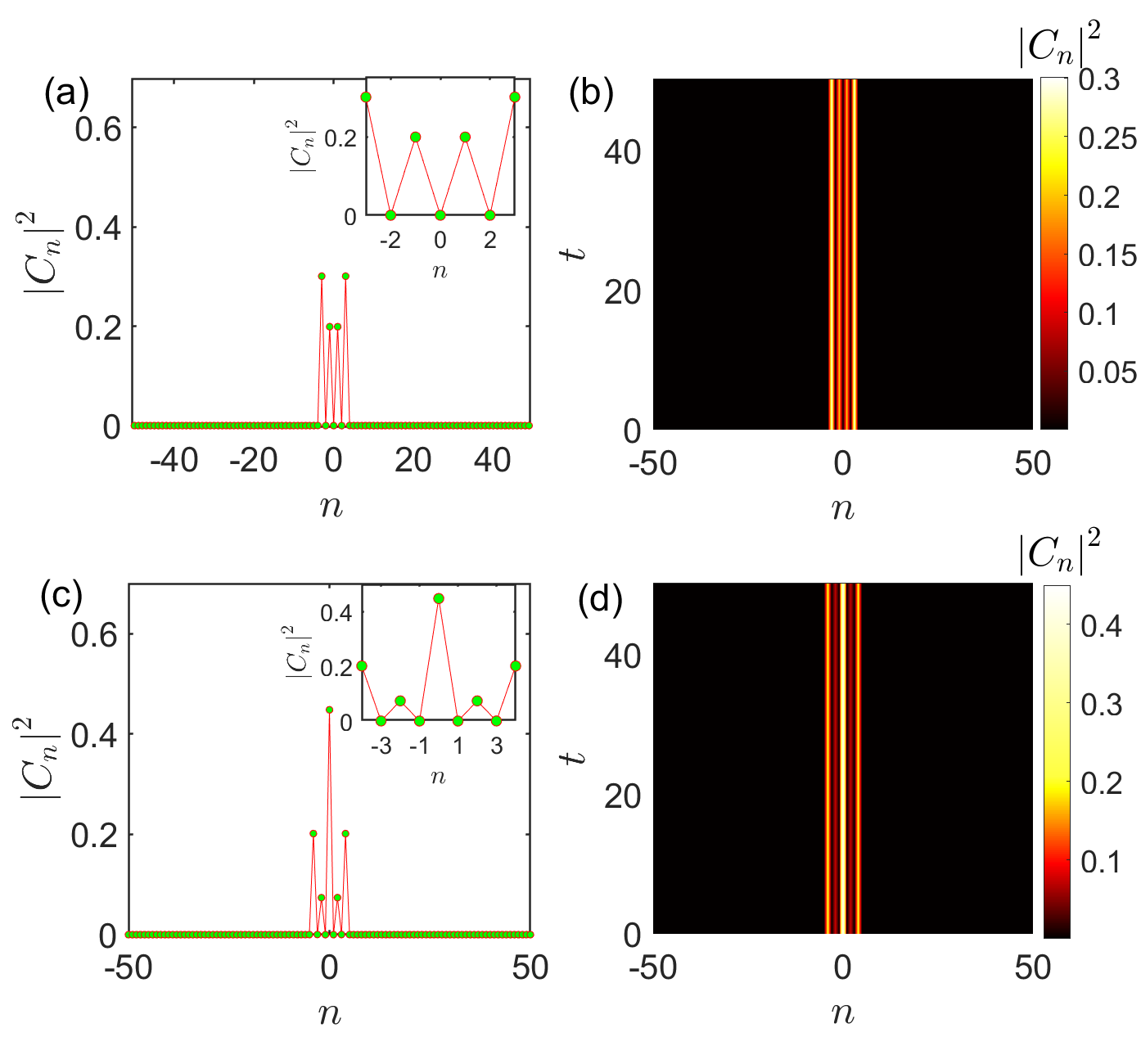}
		\caption{(a) Spatial density profile of the seven-mode dark Floquet BIC calculated from Eqs. $(\ref{con:1})$ and $(\ref{con:24})$ with $M=4$. (b) Corresponding quantum dynamics initialized with the seven-mode dark Floquet BIC state. (c) Nine-mode dark Floquet BIC spatial distribution derived from Eqs. $(\ref{con:1})$  and $(\ref{con:24})$ with $M=5$. (d) Temporal evolution of the nine-mode dark Floquet BIC. Parameters: $k=0.3$, $g=0.7k$, $a=1$, $\omega=10$, $\gamma=1$, $\Gamma=2.40509$, $N=101$.}
		\label{fig9}
	\end{figure}

  For \( M = 4 \), the parameters are configured as follows:  \( K_{-4} = K_{-3} = K_{-2} = \cdots = K_{3} = g \), and \( K_n = k \) for all other \( n \);   \( \gamma_{-2} = \gamma_{0} = \gamma_{2} = \gamma \), and \( \gamma_n = 0 \) otherwise. By tuning \( \Gamma \) to enforce \( \Theta_n = 0 \) at \( n = -4 \) and \( n =3 \) , the couplings between \( \left| -4 \right\rangle \) and \( \left| -3 \right\rangle \), and between \( \left| 3 \right\rangle \) and \( \left| 4 \right\rangle \), are eliminated. This results in a seven-state subsystem governed by the Hamiltonian:  
	\begin{equation}\label{con:25}
		\begin{split}
	\hat{H}_{\mathrm{eff}} =\begin{bmatrix}
		0 & \zeta  & 0 & 0 & 0 & 0 & 0\\
		\zeta & -i\gamma  & \zeta & 0 & 0 &0  & 0\\
		0 & \zeta & 0 & \zeta & 0 &0  & 0\\
		0 & 0 & \zeta & -i\gamma & \zeta & 0 & 0\\
		0 & 0 & 0 & \zeta & 0 & \zeta & 0\\
		0 & 0 & 0 & 0 & \zeta & -i\gamma & \zeta \\
		0 & 0 & 0 & 0 & 0 & \zeta & 0
	\end{bmatrix}.
	\end{split}
\end{equation}
The seven-state system described in Eq. $(\ref{con:25})$ also possesses a zero-energy dark state. When \( \Gamma \) is tuned to \( \Gamma = \text{2.40509} \), numerical simulations confirm the existence of a dark Floquet BIC, characterized by negligible populations in the states \( \left| -2 \right\rangle \), \( \left| 0 \right\rangle \), and \( \left| 2 \right\rangle \), as illustrated in Fig.\ \ref{fig9}(a). The evolution of this dark state remains highly stable, as demonstrated in Fig.\ \ref{fig9}(b). We further extend the system to \( M = 5 \) case to identify the nine-state dark Floquet BIC. For \( M = 5 \), tuning \( \Gamma \) to enforce \( \Theta_n = 0 \) at \( n = -5 \) and  \( n = 4 \)  (i.e., eliminating the coupling between \( \left| -5 \right\rangle \) and \( \left| -4 \right\rangle \), and between \( \left| 5 \right\rangle \) and \( \left| 4 \right\rangle \)), a nine-state subsystem is realized, as shown in Fig.\ \ref{fig9}(c). This system also hosts a zero-energy dark state, with nearly vanishing populations in the \( \left| -3 \right\rangle \), \( \left| -1 \right\rangle \), \( \left| 1 \right\rangle \), and \( \left| 3 \right\rangle \) states [Fig.\ \ref{fig9}(c)]. The dark state exhibits remarkable dynamical stability, as demonstrated by its time evolution in Fig.\ \ref{fig9}(d). Through precise engineering of Floquet parameters, we demonstrate the robust realization of stable programmable multi-mode dark BICs in a non-Hermitian lossy quantum system.

	\section{CONCLUSION}
	In summary, we have demonstrated a distinct mechanism for the generation of BICs in a dissipative Floquet system, which describes coherent particle dynamics in a driven one-dimensional tight-binding lattice. Among these Floquet BICs, we identify a novel class of dynamically stable dark Floquet BICs, which arise from the unique dark Floquet states with zero quasi-energy. By strategically amplifying the dissipation coefficient (\(\gamma\)) or driving frequency (\(\omega\)), the dark Floquet BIC exhibits negligible population leakage and robust dynamical stability, as evidenced by its nearly invariant shape during time evolution. Unlike conventional Floquet-bulk BICs \cite{36} in the nonlossy lattice system, which require fine-tuned parameters, we find that dissipation can extend the parameter ranges for the existence of BICs. These dissipation-induced BICs, which do not exist in Hermitian systems, despite being unstable, can induce total reflection of wave packets. In addition, we have discovered the existence of a nonlinear dark Floquet BIC state, which can evolve stably over a long period while maintaining its profile unchanged. In the high-frequency regime, we provide analytical insights into the existence of dark Floquet BICs using a high-frequency expansion (HFE), demonstrating their presence in the proximity of the DL regime. The interplay between periodic driving and dissipation provides a new pathway for generating stable BICs in dissipative systems. This approach permits arbitrary scalability, enabling the realization of arbitrary stable multi-mode dark Floquet BICs in a non-Hermitian system. 
	
	 On the experimental side, our theoretical model  can be fully validated with current experimental technologies. For instance, Floquet BIC states can be observed in quantum or classical simulators of the tight-binding Hamiltonian [as shown in Eq.\ (\ref{con:1})]. Specifically, experimental platforms such as ultracold atoms trapped in oscillating (or accelerating) optical lattices, or light propagation in arrays of periodically modulated (or curved) waveguides, can be utilized to simulate such dynamic behaviors. Moreover, for this periodically driven and dissipative system, quantum walk experiments offer another viable avenue for observing the stable dark Floquet BICs.
	 
	 \section{ACKNOWLEDGMENTS}
	 The work was supported by the National Natural Science Foundation of China (Grants No. 12375022 and No. 11975110), the Natural Science Foundation of Zhejiang Province (Grant No. LY21A050002), Zhejiang SciTech University Scientific Research Start-up Fund (Grant No. 20062318-Y), Jiangxi Provincial Natural Science Foundation(No. 20232BAB201008), and Education Department of Jiangxi Province (GJJ2201629).


\begin{thebibliography}{55}
		\bibitem{Neumann1929} J. V. Neumann, E. P. Wigner, $\ddot{\mathrm{U}}$ber merkwurdige diskrete Eigenwerte, Phys. Z. 30, 465–467 (1929).
		\bibitem{2} F. H. Stillinger and D. R. Herrick, Bound states in the continuum, Phys. Rev. A. 11, 446–454 (1975).
		\bibitem{3} H. Friedrich and D. Wintgen, Physical realization of bound states in the continuum, Phys. Rev. A. 31, 3964–3966 (1985).
		\bibitem{4} L. S. Cederbaum, R. S. Friedman,  V. M. Ryaboy, and N. Moiseyev, Conical Intersections and Bound Molecular States Embedded in the Continuum, Phys. Rev. Lett. 90, 013001 1–4 (2003).
		\bibitem{5} J. M. Zhang, D. Braak, and M. Kollar, Bound States in the Continuum realized in the one-dimensional Two-Particle Hubbard Model with an Impurity, Phys. Rev. Lett. 109, 116405 (2012).
		\bibitem{6} S. Longhi and  G. Della Valle, Tamm-Hubbard surface states in the continuum, J. Phys.Condens. Matter 25, 235601 (2013).
		\bibitem{7} B. N. Huang, Y. G. Ke, H. H. Zhong, Y. S. Kivshar, and C. H. Lee, Interaction-induced multiparticle bound states in the continuum, Phys. Rev. Lett. 129, 2312 (2024).
		\bibitem{8} B-J. Yang, M. S. Bahramy,and N. Nagaosa, Topological protection ofbound states against the hybridization, Nat. Com. 4, 1524 (2013).
		\bibitem{9} S. Longhi, Bound states in the continuum in a single-level Fano-Anderson model, Eur. Phys. J. B. 57, 45–51 (2007).
		\bibitem{10} D. C.Marinica, A. G. Borisov, and S. V. Shabanov Bound States in the Continuum in Photonics, Phys. Rev. Lett. 100, 183902 (2008).
		\bibitem{11} E. N. Bulgakov and A. F. Sadreev,  Bound states in the continuum in photonic waveguides inspired by defect, Phys. Rev. B. 78, 075101 (2008).
		\bibitem{12} Y. Plotnik, O. Peleg, F. Dreisow, M. Heinrich, S. Nolte, A. Szameit,  and M. Segev,  Experimental observation of optical bound states in the continuum, Phys. Rev. Lett. 107, 183901 (2011).
		\bibitem{13} K. Yamanouchi and K. Shibayama, Propagation and amplification of rayleigh waves and piezoelectric leaky surface waves in liNbO3, J. Appl. Phys. 43, 856 (1972)
		\bibitem{14} A. A. Lyapina, D. N. Maksimov, A. S. Pilipchuk, and A. F. Sadreev, Bound states in the continuum in open acoustic resonators, J. Fluid. Mech. 780, 370 (2015).
		\bibitem{15} I. Deriy, I.Tonftul, M. Peteuv, and A.bogfanov, Bound states in the continuum in compact acoustic resonators, Phys. Rev. Lett. 128, 084301 (2022).
		\bibitem{16} Z. L. Zhou, B. Jia, N. Y. Wang, X. Wang, and Y. Li, Observation of perfectly-chiral exceptional point via bound state in the continuum, Phys. Rev. Lett. 130, 116101 (2023).
		\bibitem{17} M. McIver, An example of non-uniqueness in the twodimensional linear water wave problem, J. Fluid. Mech. 315, 257 (1996).
		\bibitem{18} M. McIver, Trapped modes supported by submerged obstacles, Proc. R. Soc. A: Math. Phys. Eng. Sci. 456, 1851 (2000).
		\bibitem{19} R. Porter, Trapping of water waves by pairs of submerged cylinders, Proc. R. Soc. A: Math. Phys. Eng. Sci. 458, 607 (2002).
		\bibitem{20} S. Weimann, Y. Xu, R. Keil, A. E.Miroshnichenko, A. T$\ddot{\mathrm{u}}$nnnermann, S. Nolte, A. A.Sukhorukov, A. Szameit, and Y. S. Kivshar Compact surface fano states embedded in the continuum of waveguide arrays, Phys. Rev. Lett. 111, 240403 (2013).
		\bibitem{21} C. W. Hsu, B. Zhen, J. Lee, S-L. Chua, S. G. Johnson, J. D. Joannopoulos, and M. Solja$\check{\mathrm{c}}$i$\acute{\mathrm{c}}$, Observation of trapped light within the radiation continuum, Nat. 499, 188 (2013).
		\bibitem{22} C. W. Hsu, B. Zhen, A. Douglas, J. D. Joannopoulos, and M. Solja$\check{\mathrm{c}}$i$\acute{\mathrm{c}}$,  Bound states in the continuum, Nat. Rev. Mater. 1, 1 (2016).
     	\bibitem{23} J. Gomis-Bresco, D. Artigas, and L. Torner, Anisotropy-induced photonic bound states in the continuum, Nat. Phot. 11, 232 (2017).
		\bibitem{24} K. Koshelev, A. Bogdanov, and Y. Kivshar, Meta-optics and bound states in the continuum, Sci. Bull. 64, 836 (2019).
		\bibitem{25} E. N. Bulgakov and D. N. Maksimov, Topological Bound States in the Continuum in Arrays of Dielectric Spheres, Phys. Rev. Lett. 118, 267401 (2017).
	    \bibitem{26} B. Zhen, C. W. Hsu, L. Lu, A. D. Stone, and M. Solja$\check{\mathrm{c}}$i$\acute{\mathrm{c}}$, Topological Nature of Optical Bound States in the	Continuum, Phys. Rev. Lett. 113, 257401 (2014).
		\bibitem{27} J. H. Shirley, Solution of the Schrodinger Equation with a Hamiltonian Periodic in Time, Phys. Rev. 138, 24 (1965).
		\bibitem{28} M. S. Rudner and N. H. Lindner, Band structure engineering and 
		non-equilibrium dynamics in Floquet topological insulators, Nat. Rev. Phys. 2, 229 (2020).
		\bibitem{29} T. Oka and S. Kitamura, Floquet Engineering of Quantum Materials, Phys. Rev. Lett. 116, 250401 (2016).
		\bibitem{30} C. Gonz$\acute{\mathrm{a}}$lez-Santander, P. A. Orellana, and F. Dom$\acute{\mathrm{i}}$nguez. Adame, Bound states in the continuum driven by ac fields, Europhys. Lett. 102, 17012 (2013).
		\bibitem{31} A. Agarwala and D. Sen, Effects of local periodic driving on transport and generation of bound states, Phys. Rev. B 96, 104309 (2017).
		\bibitem{32} H. Zhong, Z. Zhou, B. Zhu, Y. Ke, and C. Lee, Floquet bound states in a driven two-particle Bose-Hubbard model with an impurity, Chin. Phys. Lett. 34, 070304 (2017).
		\bibitem{33} G. Della Valle and S. Longhi, Floquet-Hubbard bound states in the continuum, Phys. Rev. B 89, 115118 (2014).
		\bibitem{34} I. L. Garanovich, A. A. Sukhorukov, and Y. S. Kivshar, Defect-Free Surface States in Modulated Photonic Lattices, Phys. Rev. Lett. 100, 203904 (2008).
		\bibitem{35} A. Szameit, I. L. Garanovich, M. Heinrich, A. A. Sukhorukov, F. Dreisow, T. Pertsch, S. Nolte, A. T$\ddot{\mathrm{u}}$nnermann, and Y. S. Kivshar, Observation of Defect-Free Surface Modes in Optical Waveguide Arrays, Phys. Rev. Lett. 101, 203902 (2008).
		\bibitem{36} S. Longhi and G. D. Valle, Floquet bound states in the continuum, Sci. Rep. 3, 2219 (2013).
		\bibitem{37} B. Zhu, Y. G. Ke, W. J. Liu, Z. Zhou, and H. H. Zhong, Floquet-surface bound states in the continuum in a resonantly driven 1D tilted defect-free lattice, Phys. Rev. A. 102, 023303 (2020).
		\bibitem{38} Y. X. Liu and S. Chen, Fate of Two-Particle Bound States in the Continuum in Non-Hermitian Systems,  Phys. Rev. Lett. 133, 193001 (2024).
		\bibitem{39} S. Weimann, M. Kremer, Y. Plotnik, Y. Lumer, S. Nolte1, K. G. Makris, M. Segev, M. C. Rechtsman, and A. Szameit, Topologically protected bound states in photonic parity–time-symmetric crystals, Nat. Mater. 10, 1038 (2016).
		\bibitem{40} Y. V. Kartashov, V. V. Konotop, and L. Torner, Bound states in the continuum in spin-orbit-coupled atomic systems, Phys. Rev. A. 96, 033619 (2017).
	    \bibitem{41} Y. V. Kartashov, C. Mili$\acute{\mathrm{a}}$n, V. V. Konotop, and L. Torner, Bound states in the continuum in a two-dimensional PT -symmetric system, Opt. Lett. 43, 000575 (2018).
		\bibitem{42} Z. C. Hu, D. Bongiovanni, D. Juki$\acute{\mathrm{c}}$, S. Xia, D. H. Song, and J. J. Xu, Nonlinear control of photonic higher-order topological bound states in the continuum, Light. Sci. Appl. 10, 164 (2021).
		\bibitem{43} J. Doppler, A. A. Mailybaev, J. B$\ddot{\mathrm{o}}$hm, U. Kuhl, A. Girschik, F. Libisch, T. J. Milburn, P. Rabl, N. Moiseyev, and S. Rotter, Dynamically encircling an exceptional point for asymmetric mode switching, Nature (London) 537, 76 (2016).
		\bibitem{44} W. D. Heiss and G. Wunner, Chiral behaviour of the wave functions for three wave guides in the vicinity of an exceptional point of third order, Eur. Phys. J. D. 71, 312 (2017).
		\bibitem{45} H. Lignier, C. Sias, D. Ciampini, Y. Singh, A. Zenesini, O. Morsch, and E. Arimondo, Dynamical control ofmatter-wave tunneling in periodic potentials, Phys. Rev. Lett. 99, 220403 (2007).
		\bibitem{46} S. Longhi, M. Marangoni, M. Lobino, R. Ramponi, and P. Laporta, Observation of Dynamic Localization in Periodically CurvedWaveguide Arrays, Phys.  Rev. Lett. 96, 243901 (2006).
		\bibitem{47} I. L. Garanovich, S. Longhi, A. A. Sukhorukov, and Y. S. Kivshar, Light propagation and localization in modulated photonic lattices and waveguides, Phys. Rep. 518, 1–79 (2012).
		\bibitem{48} D. H. Dunlap and V. M. Kenkre, Dynamic localization of a charged particle moving under the influence of an electric field, Phys. Rev. B 34, 3625–3633
		 (1986).
        \bibitem{55} X. B. Luo, L. P. Li, L. You, and B. Wu, Coherent destruction of tunneling and dark Floquet state, New. J. Phys. 16 013007 (2014).	 
		\bibitem{54} H. Z. Wu, X. Yan, C. W. Fan, B. Y. Yang, J. P. Xiao, Z. Y. Zeng, Y. J. Chen, and X. B. Luo, Spin–orbit coupling effects on localization and correlated tunneling for two interacting bosons in a double-well potential, New. J. Phys. 26, 043020 (2024).
		 \bibitem{53} K. Bergmann, H. Theuer, and B. W. Shore, Coherent population transfer among quantum states of atoms and molecules, Rev. Mod. Phys. 70, 1003 (1998). 
		\bibitem{49} L. Wang, Q. Liu, and Y. B. Zhang,  Quantum dynamics on a lossy non-Hermitian lattice, Chin. Phys. B. 30, 020506 (2021).
		\bibitem{50} N. Goldman and J. Dalibard, Periodically Driven Quantum Systems: Effective Hamiltonians and Engineered Gauge Fields, Phys. Rev. X. 4, 031027 (2014).
		\bibitem{51} E. N. Blose, Floquet topological phase in a generalized PT-symmetric lattice, Phys. Rev. B. 102, 104303 (2020).
		\bibitem{52} A. Eckardt and E. Anisimovas, High-frequency approximation for periodically driven quantum systems from a Floquet-space perspective, New. J. Phys. 17, 093039 (2015).
	
	\end{thebibliography}
\end{document}